\documentclass[10pt]{article}
\usepackage[T1]{fontenc}
\usepackage[margin=0.9in]{geometry}
\usepackage{amsmath,amssymb,graphicx,booktabs}
\usepackage[section]{placeins}
\usepackage[colorlinks=true,urlcolor=blue,citecolor=blue,linkcolor=black]{hyperref}
\hypersetup{pdftitle={KernelMix: Bayesian kernel machine regression in Python},pdfauthor={Nilson Chapagain}}
\title{KernelMix: Bayesian kernel machine regression in Python}
\author{Nilson Chapagain\\Department of Statistics, Texas A\&M University}
\date{}
\begin{document}
\maketitle
\begin{abstract}
Exposures measured together can have nonlinear associations with an outcome, and the association with one exposure may depend on another. Bayesian kernel machine regression represents these patterns through a flexible exposure--response function. We implement this approach in KernelMix, a Python package. To handle different outcome types, KernelMix combines the exposure function with continuous, binary, count and survival models, including selected extensions for clustered observations and censored event times. For fitted models, it returns posterior summaries, uncertainty intervals, joint predictions and sampling diagnostics. Fitting and prediction use NumPy and SciPy without requiring R. To assess the implementation, we compare selected Gaussian and probit posterior summaries and conditional calculations with installed R \texttt{bkmr} 0.2.2 and check additional model components against independent mathematical calculations. A daily ozone analysis illustrates how these components form an analysis workflow.
\end{abstract}

\section{Motivation}
Exposure-mixture analyses ask how an outcome varies with several exposures measured together. A model with only separate linear exposure terms cannot represent curved associations or an association that changes with another exposure. Bayesian kernel machine regression (BKMR) replaces those terms with a joint exposure--response function, allowing these patterns without specifying each nonlinear term and interaction in advance \cite{bobb2015}.

We develop KernelMix to provide this analysis workflow in Python. The package combines model fitting with exposure contrasts, joint prediction, uncertainty summaries and sampling checks. NumPy \cite{numpy} and SciPy \cite{scipy} handle these calculations, so fitting and prediction do not require R. Our contribution is the software implementation and its documented numerical checks; the BKMR formulation is due to Bobb and colleagues.

\section{Related software}
The R package \texttt{bkmr} provides Gaussian and probit BKMR models and summaries of exposure associations \cite{bobb2018}. We use installed \texttt{bkmr} 0.2.2 as a numerical reference for selected calculations in these models. Constructed parameter states check conditional calculations. We also compare fitted posterior summaries under matched data and priors, allowing for Monte Carlo error from finite sampling.

KernelMix also supports count and survival outcomes, additional cluster and residual models, and alternative kernel representations. These options have separate likelihoods, priors and prediction rules, so agreement with the Gaussian and probit reference does not validate them.

\section{Methods and implementation}
We first describe how the joint exposure function enters a model for a continuous outcome. Following Bobb et al.\ \cite{bobb2015}, let $Y_i$ be the outcome for observation $i$, $\mathbf z_i$ its $M$ exposures, and $\mathbf x_i$ its adjustment covariates. We write the joint exposure contribution as $h(\mathbf z_i)$ and retain linear adjustment through coefficients $\boldsymbol\beta$. Together, these terms give the Gaussian model
\[
Y_i=h(\mathbf z_i)+\mathbf x_i^{\mathsf T}\boldsymbol\beta+\epsilon_i,\qquad
\epsilon_i\overset{\mathrm{iid}}{\sim}N(0,\sigma^2).
\]
The function $h$ contains the nonlinear associations and interactions, while $\epsilon_i$ represents independent error with variance $\sigma^2$.

Fitting $h$ requires a way to relate its values at different exposure combinations. We specify this relationship through a Gaussian process (GP) prior. A prior describes parameter or function values before using the outcome data; the posterior updates that distribution using the data. For $n$ observations, write $\mathbf h=(h(\mathbf z_1),\ldots,h(\mathbf z_n))^{\mathsf T}$. The GP prior relates these values through the Gaussian kernel \cite{bobb2015,gpml},
\[
\mathbf h\sim N(\mathbf0,\tau\mathbf K_{Z,\mathbf r}),\qquad
K(\mathbf z,\mathbf z';\mathbf r)=
\exp\!\left\{-\sum_{m=1}^{M}r_m(z_m-z'_m)^2\right\}.
\]
The kernel $K$ describes similarity between exposure combinations, while $\tau$ controls variation in $h$. Each $r_m\geq0$ controls how exposure $m$ affects that similarity. Setting $r_m=0$ removes exposure $m$ from $h$, which permits exposure selection. Group selection restricts each included group to at most one exposure \cite{bobb2015}. Inclusion probabilities summarize selection under the fitted model; neither they nor $r_m$ measure causal importance.

The Gaussian model above combines the exposure function and linear adjustment with a continuous outcome. To handle other outcomes, we retain those two contributions but change their link to the outcome distribution. For binary outcomes, the probit model gives outcome one when an underlying normal variable exceeds zero \cite{bobb2018}. Poisson and negative-binomial models describe counts \cite{standocs}. For event times, a piecewise-exponential model uses a baseline hazard that is constant within each time interval \cite{inla}. A hazard is an instantaneous event rate among individuals still at risk \cite{standocs}.

Event-time fitting also depends on what is observed about each event. Survival records can specify entry after time zero, competing event causes, or an event known only to occur within an interval. Supplied covariate histories remain constant within each recorded risk interval. With delayed entry, inference concerns the included cohort and does not recover the distribution of unobserved risk before entry.

Beyond the outcome type, repeated observations may share variation within a cluster. Random intercepts and slopes represent this variation through shared coefficients \cite{bobb2018,nlme}. For Gaussian outcomes, residual models can separately specify unequal error variances or correlation within clusters \cite{nlme}.

These outcome and dependence options are distinct from the kernel that relates exposure combinations. The Gaussian kernel can be replaced by Mat\'ern $3/2$ or rational-quadratic kernels \cite{gpml}. To reduce the number of GP coordinates, we also provide predictive-process and paired random-Fourier representations \cite{pp,rff}. This reduction changes the model approximation: the predictive process omits variation remaining after conditioning on knots, and random-Fourier inference conditions on the sampled features. Large residual-correlation blocks and joint prediction requests can still require dense matrices.

These model choices also affect sampling and prior conventions. The default Gaussian sampler averages over $\mathbf h$ analytically during parameter sampling; default Gaussian and probit fits store $\lambda=\tau/\sigma^2$, with $\sigma^2=1$ for probit. Count, survival and structured fits instead store the direct amplitude $\tau$ as \texttt{lambda} and use proper normal priors for adjustment coefficients. The Gaussian-prior blocks in these fits use elliptical slice sampling \cite{murray}. Priors therefore need to be compared on their stated parameter scales. The manual specifies the inputs, priors and prediction rules for each option.

\section{Software checks}
We check fitted posterior summaries against installed R \texttt{bkmr} 0.2.2 on four fixed datasets: Gaussian and probit models, each with and without random intercepts. Both implementations use identical observations, exposure standardization, adjustment design and priors, with exposure selection disabled. Four chains per implementation provide means and 10th, 50th and 90th posterior percentiles for parameters, three values of $h$ and one exposure contrast. Each R chain retains 120,000 draws after 3,000 warmup transitions; each Python chain retains 60,000. All 160 checked summaries pass the fixed diagnostic, precision and discrepancy rules. We require rank-normalized/folded split $\widehat R\leq1.01$, bulk/tail effective sample sizes at least 400 and Monte Carlo standard errors at most 2\% of the pooled posterior standard deviation. The manual gives the complete comparison rules and results; these numerical screens do not establish posterior equivalence or general convergence. Constructed-state checks additionally compare conditional means, covariances, inclusion probabilities and prior densities.

For models without an R reference, we compare selected posterior moments with independent numerical integration. Three fixed cases cover negative-binomial counts, right-censored survival and a Gaussian model with random intercepts/slopes and AR1 residual correlation. All use a constant exposure function, with its GP amplitude sampled. The structured case fixes random-coefficient covariance and residual correlation. All 12 checked moments and four selected covariances pass their error and precision rules. These small cases assess the stated posterior quantities, not every structured option or general convergence.

An installed wheel passes 512 tests in each of two native macOS environments: Python 3.14 with current dependencies and Python 3.10 with minimum dependencies. Configured hosted Linux and Windows jobs stop before runner startup, so Windows and hosted Linux compatibility remain unverified.

Numerical agreement is distinct from reproducing the original BKMR studies. The original simulations used 300 datasets per scenario with $n=100$ and resampled empirical exposure rows; the applications analyzed 375 children and 142 exposure records from 13 dogs \cite{bobb2015}. The original Bangladesh, Boston and dog datasets and sufficient fitting settings were not recovered from the inspected open assets. Gaussian-generated exposures with a similar covariance would define a different experiment. We therefore use the observed-data example below to demonstrate the workflow, without treating it as original-study reproduction or a study of interval coverage.

\section{Observed-data example}
\subsection{Data}
We use the public \texttt{airquality} data to examine how daily ozone varies with solar radiation, wind and temperature, after adjustment for calendar month \cite{rcore,chambers}. The data contain 153 New York daily records from May--September 1973. Ozone and weather measurements come from different stations and times, so they do not describe a person's joint exposures.

\begin{figure}[!htbp]
\centering
\includegraphics[width=\linewidth]{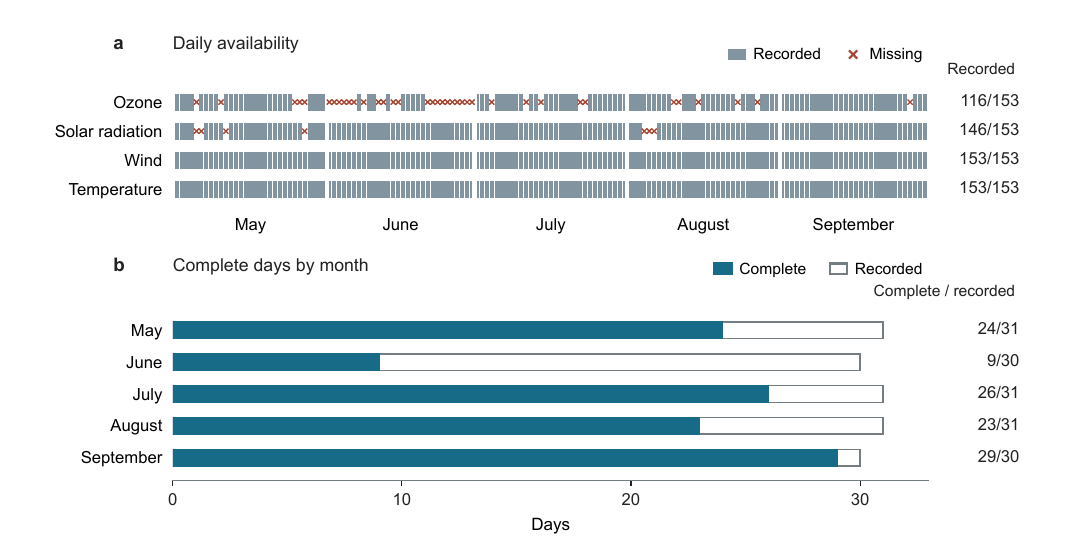}
\caption{Measurement availability among 153 daily records, May--September 1973. (a) Each cell represents one day: filled cells indicate recorded measurements and crosses indicate missing values. Fractions report recorded measurements out of 153 days. (b) Filled bars show complete days within outlines for all recorded days; fractions give complete/recorded counts. The analysis uses 111 complete days. Missing values are neither imputed nor treated as zero.}
\label{fig:availability}
\end{figure}

\begin{figure}[!htbp]
\centering
\includegraphics[width=\linewidth]{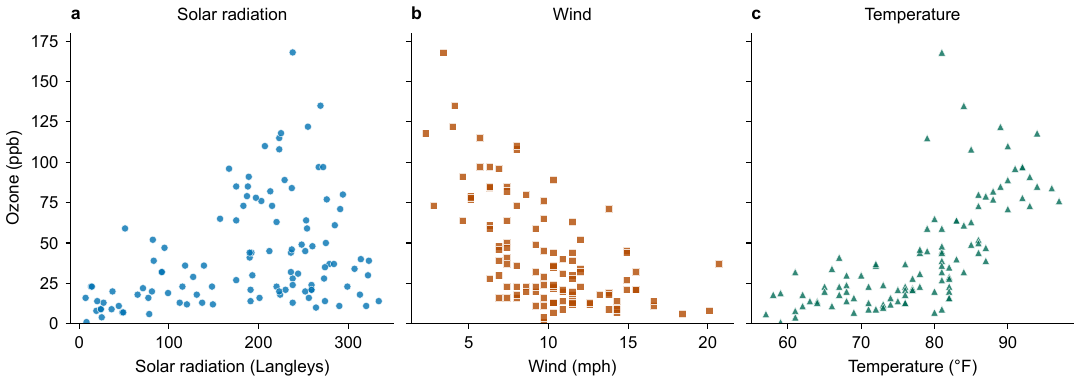}
\caption{Ozone against solar radiation, wind and temperature among the 111 complete days. Each point represents one day. The plots show unadjusted observations, without holding other weather variables or month fixed.}
\label{fig:observed}
\end{figure}

Before fitting, we identify the days with all required measurements: 111 are retained and 42 are excluded. There are 37 missing ozone values and seven missing solar-radiation values, with overlap. Figure~\ref{fig:availability} shows that this exclusion is uneven across months: June retains 9 of 30 days, compared with 29 of 30 in September. For these retained days, Figure~\ref{fig:observed} shows ozone against each weather measurement. These pairwise plots leave the other weather variables and month unadjusted.

\FloatBarrier

\subsection{Model and fitted associations}
To estimate the weather associations together and adjust for month, we fit the Gaussian model to $Y_i=\log(\mathrm{Ozone}_i)$, with solar radiation, wind and temperature in $\mathbf z_i$. The adjustment vector $\mathbf x_i$ contains an intercept and four indicators for June through September, with May as the reference. We standardize exposures using the retained days. This month adjustment represents broad seasonal differences; the independent-error assumption does not account for dependence between consecutive days.

We use independent $r_m\sim\mathrm{Gamma}(1,1)$ priors, an inverse-gamma$(2,1)$ prior for $\sigma^2$, $\lambda\sim\mathrm{Gamma}(1,0.1)$ and the default flat prior for $\boldsymbol\beta$. Gamma parameters are shape and rate. Exposure selection is disabled, so this analysis estimates the joint exposure function without selecting a subset of weather variables.

With the model and priors specified, we ran four chains sequentially with seed 73. Each used 5,000 discarded warmup iterations followed by 20,000 saved draws. An initial run with 4,000 saved draws had maximum $\widehat R=1.0142$, after which we extended sampling without changing the model, data or priors. Curve and fitted-residual summaries use 400 evenly spaced draws per chain. Fitting took 72.6 seconds in the recorded local environment.

In the longer run, the maximum $\widehat R$ was 1.0053, the minimum bulk effective sample size was 885 and the minimum tail effective sample size was 1,193. These summaries concern ten individual parameters; the full diagnostics are shown below.

\begin{figure}[!htbp]
\centering
\includegraphics[width=\linewidth]{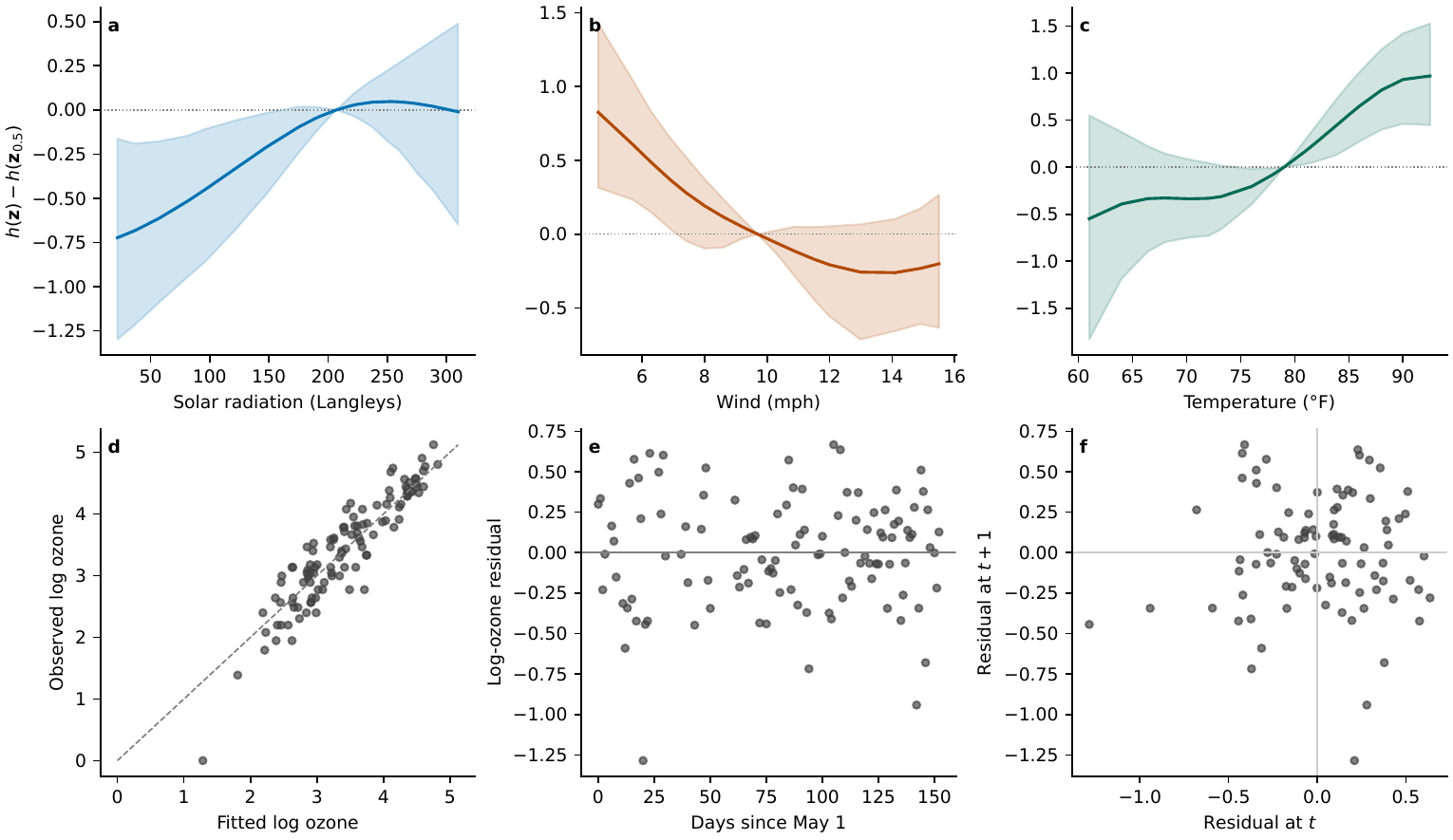}
\caption{Fitted exposure associations and residuals ($n=111$). (a--c) Posterior means and 95\% pointwise credible intervals for $h(\mathbf z)-h(\mathbf z_{0.5})$, with other exposures fixed at their medians and $\mathbf z_{0.5}$ the all-median profile. (d--f) Fitted versus observed log ozone, residuals by date, and residual pairs one calendar day apart. Missing dates are not connected. Intervals assume independent Gaussian working errors.}
\label{fig:ozone}
\end{figure}

Figure~\ref{fig:ozone} separates each fitted weather association by varying one exposure while fixing the others at their medians. The posterior mean curve generally rises with temperature and falls with wind; the solar-radiation curve is nonlinear. These are fitted conditional associations, with pointwise uncertainty under the independent-error model.

The displayed contrast is $h(\mathbf z)-h(\mathbf z_{0.5})$, where $\mathbf z_{0.5}$ is the vector of exposure medians. On the log-ozone scale, exponentiating this contrast gives a ratio of fitted ozone medians at the two exposure combinations, with month held fixed. It does not give a difference in arithmetic means.

To inspect the variation remaining after fitting, we examine the fitted-versus-observed and residual panels. These use the same days used for fitting, so they assess the fitted sample rather than prediction on new data. Among 92 residual pairs from consecutive calendar days, the sample correlation is 0.0136; this small correlation does not establish independent errors.

\begin{figure}[!htbp]
\centering
\includegraphics[width=\linewidth]{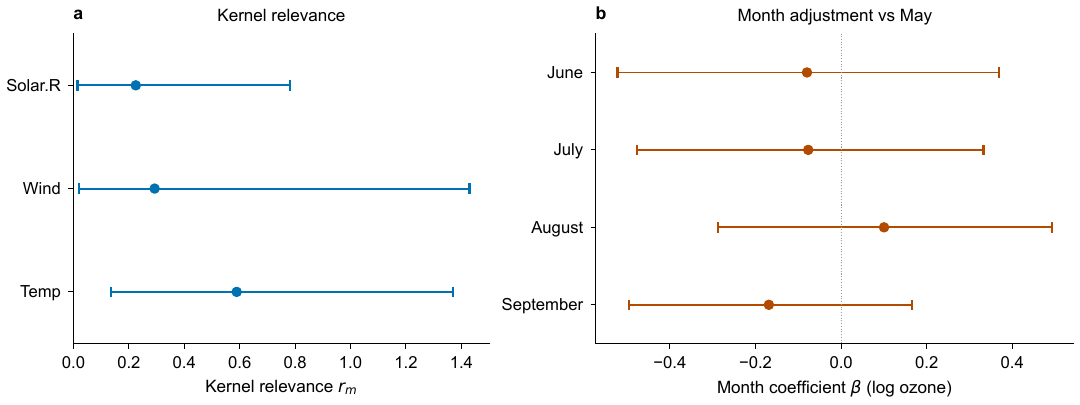}
\caption{Posterior means and 95\% equal-tailed credible intervals for (a) kernel relevance parameters $r_m$ and (b) month coefficients in $\boldsymbol\beta$, relative to May, on the log-ozone scale. The intercept is omitted. The panels have different scales; kernel relevance does not measure causal importance.}
\label{fig:parameters}
\end{figure}

The fitted model separates the exposure function from month adjustment. Figure~\ref{fig:parameters} reports their parameters separately, distinguishing kernel relevance $r_m$ from month coefficients $\boldsymbol\beta$. Their different roles require separate axes. Each 95\% credible interval contains 95\% of the posterior probability under the fitted model. Because exposure selection is disabled, positive $r_m$ values describe the fitted kernel and do not establish variable-selection evidence.

\FloatBarrier

\subsection{Sampling checks and exposure support}
\begin{figure}[!htbp]
\centering
\includegraphics[width=\linewidth]{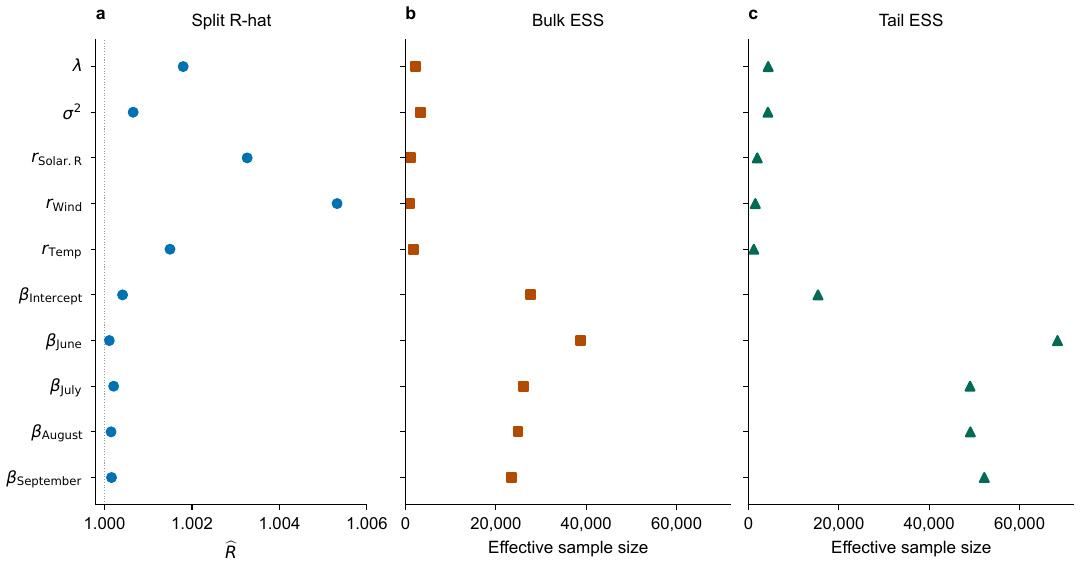}
\caption{Sampling diagnostics for ten parameters from four chains, each with 20,000 saved draws. (a) Rank-normalized/folded split $\widehat R$. (b) Bulk effective sample size. (c) Tail effective sample size. These diagnostics assess sampling, not the model assumptions.}
\label{fig:diagnostics}
\end{figure}

The preceding intervals depend on adequate posterior sampling. We assess sampling with split $\widehat R$, which compares chains, and effective sample sizes, which account for dependence between draws \cite{vehtari} (Figure~\ref{fig:diagnostics}). Bulk effective sample size concerns the main posterior distribution, while tail effective sample size concerns its tails. These checks cover ten individual parameters; favorable values do not prove convergence of every exposure curve or validate the model assumptions.

\begin{figure}[!htbp]
\centering
\includegraphics[width=\linewidth]{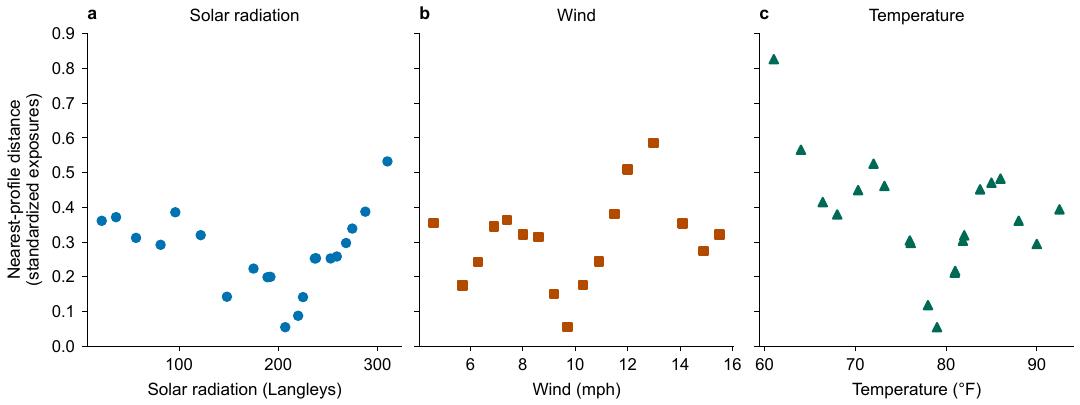}
\caption{Nearest observed-combination distance after exposure standardization. Each plotted value is combined with medians of the other exposures. The grids contain 21 solar-radiation, 17 wind and 21 temperature values. Distance measures proximity to the complete observed data, without a model-validity threshold.}
\label{fig:support}
\end{figure}

Adequate sampling also does not ensure that the plotted exposure combinations occur in the data. Each curve uses the exposure's 5th--95th percentiles, with duplicate values removed, while fixing the other exposures at their medians. This construction can produce combinations far from observed rows. We therefore report the distance to the nearest complete exposure combination after training standardization (Figure~\ref{fig:support}). Larger distances indicate less nearby data; no cutoff is used to declare a prediction reliable.

\FloatBarrier

\section{Limits and reproducibility}
The ozone curves describe conditional associations under the specified priors and independent Gaussian errors. Month adjustment does not model daily dependence, and complete-case analysis may introduce selection bias \cite{sterne}. Sampling diagnostics address neither issue, so the analysis supports no causal or forecasting claim. The numerical checks also leave repeated-dataset bias, interval coverage, original-study reproduction and large-data performance unresolved.

To make these calculations and the ozone workflow reproducible, the repository at \url{https://github.com/nilson01/kernelmix} includes installation instructions, a function reference, tests, source data and analysis scripts. Saved tables record the data hash, fitting settings, environment and diagnostics. Figure scripts use Matplotlib \cite{matplotlib}. The software is distributed under GPL-2.0; the R reference and data sources are credited separately.

\section*{Acknowledgements}
Computations were performed on the Arseven Computing Cluster, Department of
Statistics, Texas A\&M University.

This work received no grant funding.

\end{document}